\documentclass[prl,aps,twocolumn,10pt,showpacs,groupedaddress,superscriptaddress,floatfix,notitlepage]{revtex4-1}
\usepackage{amsmath,amsfonts,amssymb,graphics,graphicx,epsfig,color,times}
\usepackage[utf8x]{inputenc}
\usepackage{color}
\usepackage{bbm, dsfont} 
\usepackage{subfigure}
\usepackage{hyperref}
\usepackage{mathrsfs}
\usepackage{verbatim}
\begin{document}
\title{Steady-state Phase Diagram of a Weakly Driven Chiral-coupled Atomic Chain}
\author{H. H. Jen}
\email{sappyjen@gmail.com}
\affiliation{Institute of Physics, Academia Sinica, Taipei 11529, Taiwan}

\date{\today}
\renewcommand{\r}{\mathbf{r}}
\newcommand{\f}{\mathbf{f}}
\renewcommand{\k}{\mathbf{k}}
\def\p{\mathbf{p}}
\def\q{\mathbf{q}}
\def\bea{\begin{eqnarray}}
\def\eea{\end{eqnarray}}
\def\ba{\begin{array}}
\def\ea{\end{array}}
\def\bdm{\begin{displaymath}}
\def\edm{\end{displaymath}}
\def\red{\color{red}}
\pacs{}
\begin{abstract}
A chiral-coupled atomic chain of two-level quantum emitters allows strong resonant dipole-dipole interactions, which enables significant collective couplings between every other emitters. We numerically obtain the steady-state phase diagram of such system under weak excitations, where interaction-driven states of crystalline orders, edge or hole excitations, and dichotomy of chiral flow are identified. We distinguish these phases by participation ratios and structure factors, and find two critical points which relate to decoherence-free subradiant sectors of the system. We further investigate the transport of excitations and emergence of crystalline orders under spatially-varying excitation detunings, and present non-ergodic butterfly-like system dynamics in the phase of extended hole excitations with a signature of persistent subharmonic oscillations. Our results demonstrate the interaction-induced quantum phases of matter with chiral couplings, and pave the way toward simulations of many-body states in nonreciprocal quantum optical systems.
\end{abstract}
\maketitle
{\it Introduction.--}A chiral-coupled atomic system \cite{Lodahl2017, Chang2018} from an atom-fiber \cite{Mitsch2014} or an atom-waveguide \cite{Luxmoore2013, Sollner2015} interface presents the capability to engineer the directionality of light transmissions. This leads to a broken time-reversal symmetry of light-matter couplings, and results in nonreciprocal decay channels. Unidirectional coupling \cite{Arcari2014} can therefore be enabled by spin-momentum locking \cite{Bliokh2014, Bliokh2015}, where light propagation highly correlates to its transverse spin angular momentum. Such one-dimensional (1D) nanophotonics system has been studied to create mesoscopic quantum correlations \cite{Tudela2013} and quantum spin dimers \cite{Stannigel2012, Ramos2014, Pichler2015}, to enable simulations of long-range quantum magnetism \cite{Hung2016}, and to manifest emerging universal dynamics \cite{Kumlin2018} and strong photon-photon correlations \cite{Mahmoodian2018}. These compelling predictions rely on the emergence of infinite-range resonant dipole-dipole interactions (RDDI) \cite{Solano2017} in 1D system-reservoir interactions, in huge contrast to the ones in free-space, which decrease with an inter-atomic distance in the long range \cite{Lehmberg1970}. 

The nonreciprocal decay channels of chiral-coupled systems can be tuned by external magnetic fields \cite{Mitsch2014, Luxmoore2013, Sollner2015}, such that the amount of light transmissions in the allowed direction can be controlled by the internal states of quantum emitters \cite{Mitsch2014}. This can be attributed to reservoir engineering, which has spurred many interesting studies of non-equilibrium phase transitions in driven-dissipative quantum systems \cite{Diehl2008, Kraus2008, Verstraete2009}. In such open systems, self-organized supersolid phase \cite{Baumann2010} of Bose–Einstein condensate can be realized by coupling to an optical cavity, and exotic spin phases and multipartite entangled states respectively can be dissipatively prepared in laser-excited Rydberg atoms \cite{Weimer2010} and ions \cite{Barreiro2011}. Under engineered dissipations, Majorana edge modes as topological states of matter \cite{Diehl2011, Bardyn2013} and critical phenomena at steady-state phase transitions \cite{Honing2012} are also predicted in lattice fermions. These quantum phases of matter under non-equilibrium phase transitions show the potential to explore dynamical phases driven by competing dissipation and interaction strengths \cite{Diehl2010}. 

In this Letter, we consider two-level quantum emitters coupled to a nanofiber or waveguide with equal inter-atomic distances. We obtain the steady-state phase diagram of such chiral-coupled atomic chain in the low saturation limit, determined by two competing parameters of directionality and dipole-dipole interaction strengths. The interaction-driven phases include the states with extended distributions (ETD), crystalline orders (CO), bi-edge/hole excitations (BEE/BHE), and of chiral-flow dichotomy (CFD), which we classify by participation ratios and structure factors. Two critical points are also located, where time to reach steady states is longer than a power-law dependence of system sizes. This critically slow equilibrium of system dynamics relates to the decoherence-free sectors of the eigen-spectrum. We further explore the possibility to relocate the atomic excitations via spatially-varying field detunings or excitation directions. Finally, in the phases of BEE and BHE, non-ergodic signatures of subharmonic oscillations emerge, and we specifically present a butterfly-like system dynamics as an example. The steady-state phases and their dynamical evolutions investigated here present a distinct interplay between RDDI and directionality of a chiral-coupled atomic chain, which give insights to preparations and simulations of many-body states in nonreciprocal quantum optical systems.

\begin{figure*}[t]
\centering
  \begin{tabular}{cc}

    \includegraphics[width=11.5cm,height=8.5cm]{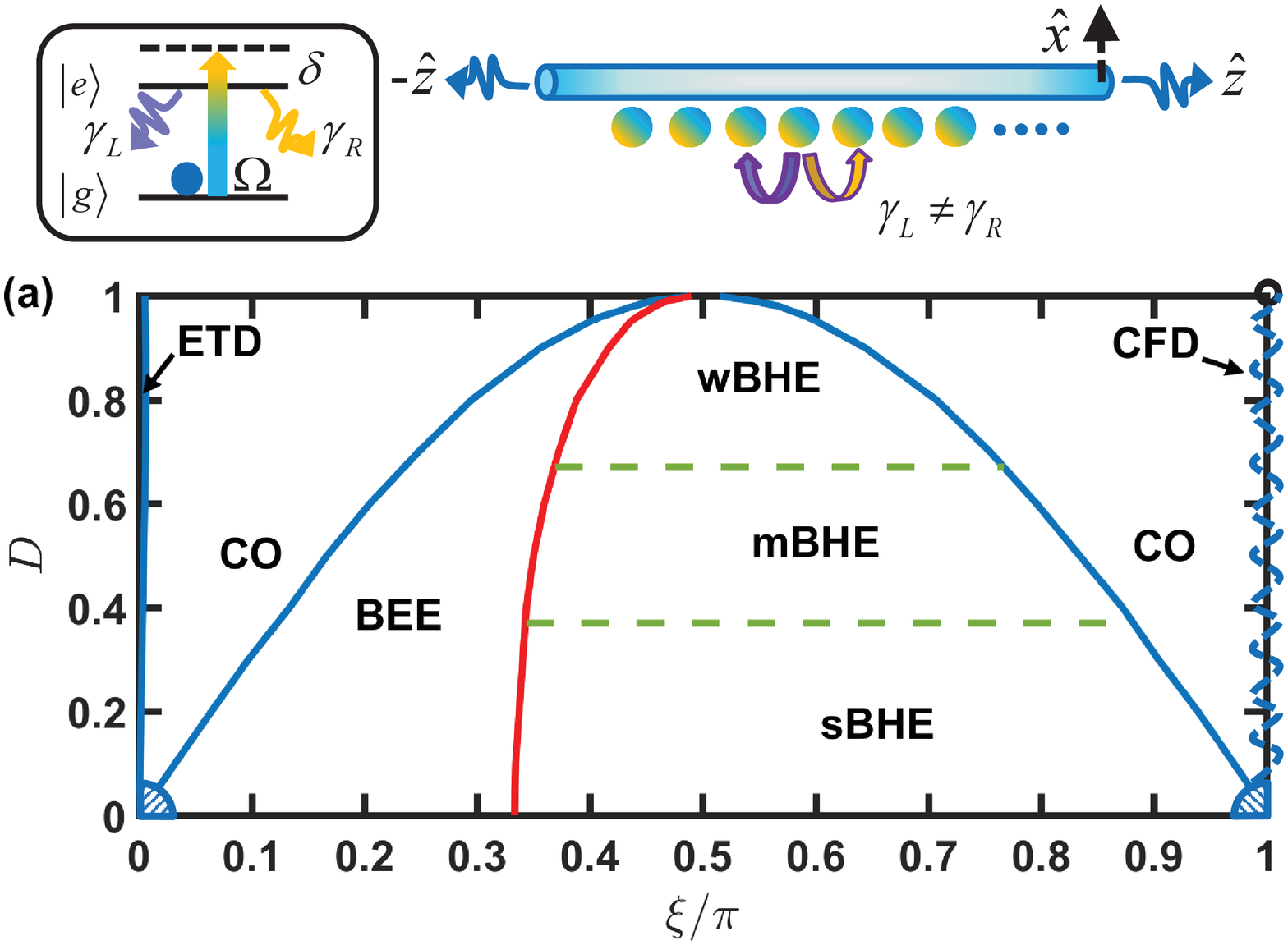}&

    \includegraphics[width=14.5cm,height=7cm]{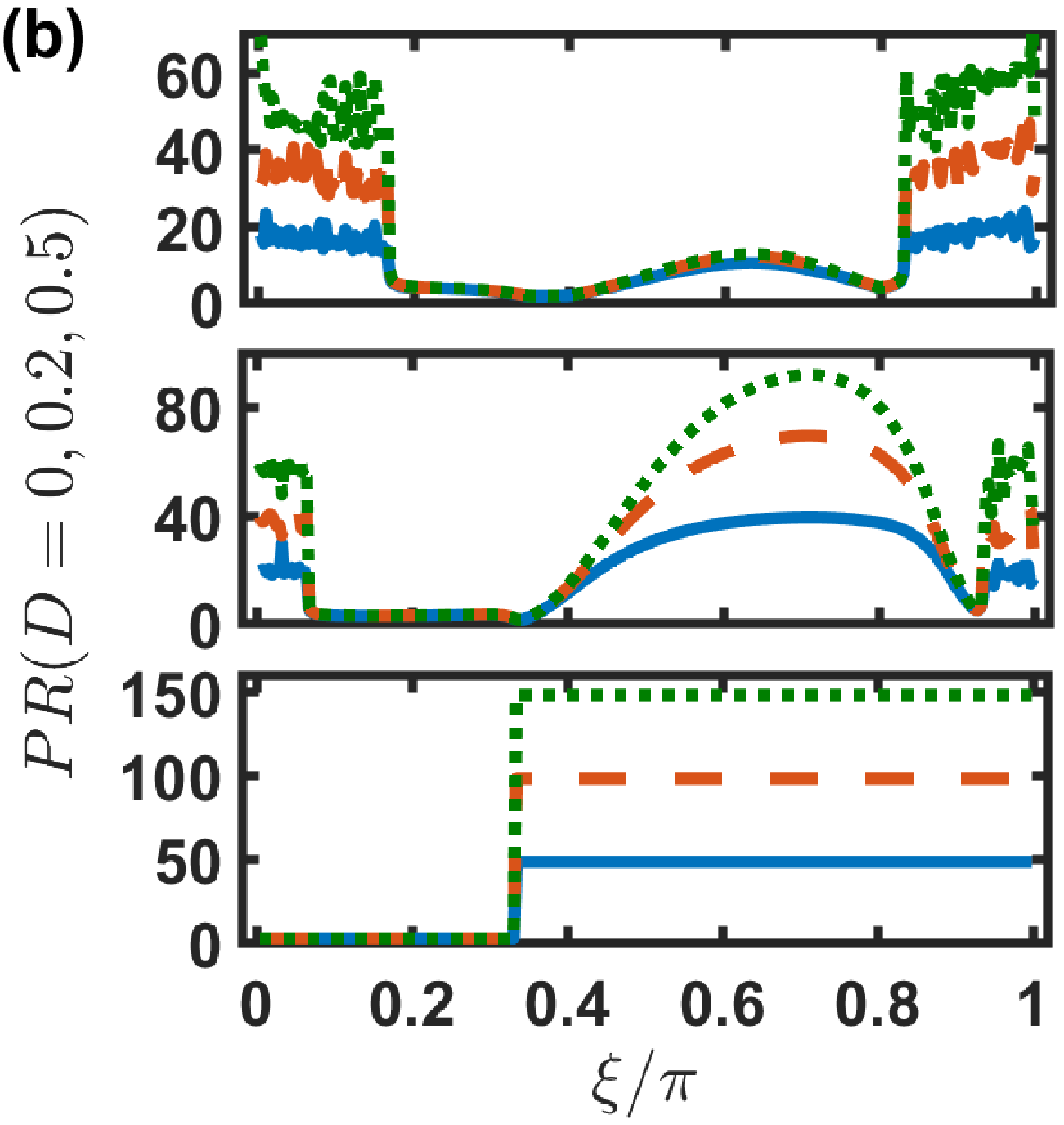}

  \end{tabular}
	\caption{Steady-state phase diagram of a weakly driven chiral-coupled atomic chain. The chiral-coupled system consists of two-level quantum emitters with nonreciprocal decay rates of $\gamma_L\neq \gamma_R$, mediated by a nanofiber or waveguide. (a) In the steady state, the phases of extended distributions (ETD), finite crystalline orders (CO), bi-edge/hole excitations (BEE/BHE), and chiral-flow dichotomy (CFD, wavy dashes) are identified under parameter spaces of directionality factor $D$ and dimensionless inter-atomic distance $\xi$. For different scalings of participation ratios on the number of atoms ($N^{\alpha}$), we further separate the BHE phase into strong (s), moderate (m), and weak (w) regimes with $\alpha>0.5$ (below $D\approx 0.37$), $0.1<\alpha<0.5$, and $\alpha<0.1$ (above $D\approx 0.67$), respectively. Shaded regions at two lower corners of the diagram represent the critical regimes. (b) Participation ratios (PR) at different horizontal cuts of the diagram (increasing $D$ from the lower to upper panels), showing abrupt changes crossing the CO phases, with $N$ $=$ $50$ (solid line), $100$ (dashes), and $150$ (dots).}\label{fig1}
\end{figure*}
{\it Model.--}We consider a generic driven-dissipative model in Lindblad forms for a 1D chiral-coupled atomic chain \cite{Pichler2015}, 
\bea
\frac{d \rho}{dt}=-\frac{i}{\hbar}[H_S+H_L+H_R,\rho]+\mathcal{L}_L[ \rho]+\mathcal{L}_R[\rho].\label{rho}
\eea
The external and coherent field excitation is denoted as 
\bea
H_S =\hbar\sum_{\mu=1}^N\Omega_\mu\left(\sigma_\mu+\sigma_\mu^\dag\right)-\hbar\sum_{\mu=1}^N\delta_\mu\sigma_\mu^\dag\sigma_\mu,
\eea
which drives $N$ two-level quantum emitters ($|g\rangle$ and $|e\rangle$ being the ground and excited state respectively) with spatially dependent Rabi frequencies $\Omega_\mu$ and detunings $\delta_\mu$, and dipole operators are $\sigma_\mu^\dag$ $\equiv$ $|e\rangle_\mu\langle g|$ with $\sigma_\mu$ $=$ $(\sigma_\mu^\dag)^\dag$. The coherent parts of RDDI are 
\bea
H_{L(R)} = -i\hbar\frac{\gamma_{L(R)}}{2} \sum_{\mu<(>)\nu}^N\left(e^{ik_s|x_\mu-x_\nu|} \sigma_\mu^\dag\sigma_\nu-\textrm{H.c.}\right),
\eea
and the dissipative ones in Lindblad forms are 
\bea
\mathcal{L}_{L(R)}[\rho]=&&-\frac{\gamma_{L(R)}}{2} \sum_{\mu,\nu}^N e^{\mp ik_s(x_\mu-x_\nu)} \left(\sigma_\mu^\dag \sigma_\nu \rho + \rho \sigma_\mu^\dag\sigma_\nu \right.\nonumber\\
&&\left.-2\sigma_\nu \rho\sigma_\mu^\dag\right).
\eea
They determine the collective energy shifts and decay rates, respectively, which mediate the whole system in infinite-range of sinusoidal forms. The subscripts $L(R)$ label the left(right)-propagating decay channels, and $k_s=2\pi/\lambda$ denotes the wave vector for the transition wavelength $\lambda$.  

We proceed to solve for the steady-state solutions of Eq. (\ref{rho}) in the low saturation limit, which truncates the hierarchy-coupled equations \cite{SM, note1} to self-consistently coupled dipole operators. We obtain steady-state $\vec\sigma^{(s)}$ satisfying $\dot{\sigma}_\mu=0$, 
\bea
\vec\sigma^{(s)}=i\Omega M^{-1} e^{ik_s\cos\theta_s\vec r},
\eea
where a uniform $\theta_s$ with small Rabi frequency $\Omega$ represents the excitation angle to the alignment of the chain, and $\vec r$ denotes the atomic distributions. In the coupling matrix $M$, asymmetry of off-diagonal matrix elements arises due to unequal $\gamma_{L(R)}$, and they are 
\bea
M_{\mu,\nu}=\left\{\begin{array}{lr}
    -\gamma_Le^{ik_s|r_{\mu,\nu}|},~\mu<\nu\\
		i\delta_\mu-\frac{\gamma_L+\gamma_R}{2},~\mu=\nu\\
		-\gamma_Re^{ik_s|r_{\mu,\nu}|},~\mu>\nu
\end{array}\right.,\label{M}
\eea
where $r_{\mu,\nu}$ $=$ $r_\mu-r_\nu$. From Eq. (\ref{M}), we determine the interaction-driven quantum phases of matter which mainly rely on the interplay between the chirality of chiral-coupled systems and RDDI determined by inter-atomic distances. The external spatially-varying excitation detunings and excitation angles play extra roles in relocating the atomic excitations and imprinting extra phases on the atoms, which we investigate below. 


{\it Steady-state phase diagram.--}We use the directionality factor $D$ $\equiv$ $(\gamma_R-\gamma_L)/\gamma$ \cite{Mitsch2014} to quantify the degree of light transmissions of the system with a normalized decay channel $\gamma_R$ $+$ $\gamma_L$ $=$ $\gamma$. $D$ $=$ $\pm 1$ and $0$ present the unidirectional \cite{Stannigel2012, Gardiner1993, Carmichael1993} and reciprocal couplings respectively. The RDDI strength can be quantified by $\xi$ $\equiv$ $k_s r_{\mu+1,\mu}$, where $\xi$ $=$ $0$ or $2\pi$ represents the strong coupling regime with vanishing dispersions that cannot be achieved in free space \cite{Lehmberg1970}. In Fig. \ref{fig1}, we numerically obtain the steady-state phase diagram of a weakly driven chiral-coupled atomic chain at $\theta_s$ $=$ $\pi/2$. The states with CO are determined by finite structure factors $S(k)=\sum_{j,m}^N e^{ik(j-m)}\tilde P_j\tilde P_m$ \cite{SM} with normalized excitation populations $\tilde P_j$ $=$ $\sigma_{j}^{(s)\dag}\sigma_{j}^{(s)}/(\sum_{j=1}^N \sigma_j^{(s)\dag}\sigma_j^{(s)})$. This phase boundary extends symmetrically from $\xi=\pi/2$ toward two critical points at the lower corners of the diagram, which also coincides with abrupt changes of participation ratio $PR$ $\equiv$ $(\sum_{j=1}^N \Delta \tilde P_j)^2/\sum_{j=1}^N(\Delta \tilde P_j)^2$ \cite{Murphy2011}, where $\Delta\tilde P_j$ $=$ $|\tilde P_j - N^{-1}|\Theta(\tilde P_j - N^{-1})$ with the Heaviside step function $\Theta$ evaluates the state variations from an ETD phase (narrow region close to $\xi$ $=$ $0$) of uniform distributions $N^{-1}$. As shown in Fig. \ref{fig1}(b), sharp rises of $PR$ emerge when $\xi$ crosses extended CO or BHE ($\tilde P_{1(N)}$ $<$ $N^{-1}$) phases. This is in huge contrast to the localized BEE phase ($\tilde P_{1(N)}$ $>$ $N^{-1}$), where $PR$ stays constantly small as system size increases. At $D$ $=$ $0$, the phase boundary between BEE and BHE starts from $\xi$ $=$ $\pi/3$ \cite{SM} and collapses to $\pi/2$ as $D$ increases. For an increasing $D$, $PR$ of BHE phase suppresses, and we further distinguish it by different scalings of system sizes \cite{SM}. The strong (s) BHE region suggests a significant extended state distributions with hole excitations at both ends of the chain, which endows the emergence of persistent subharmonic dynamics we will present below.

\begin{figure}[t]
\centering
\includegraphics[width=8.5cm,height=4.5cm]{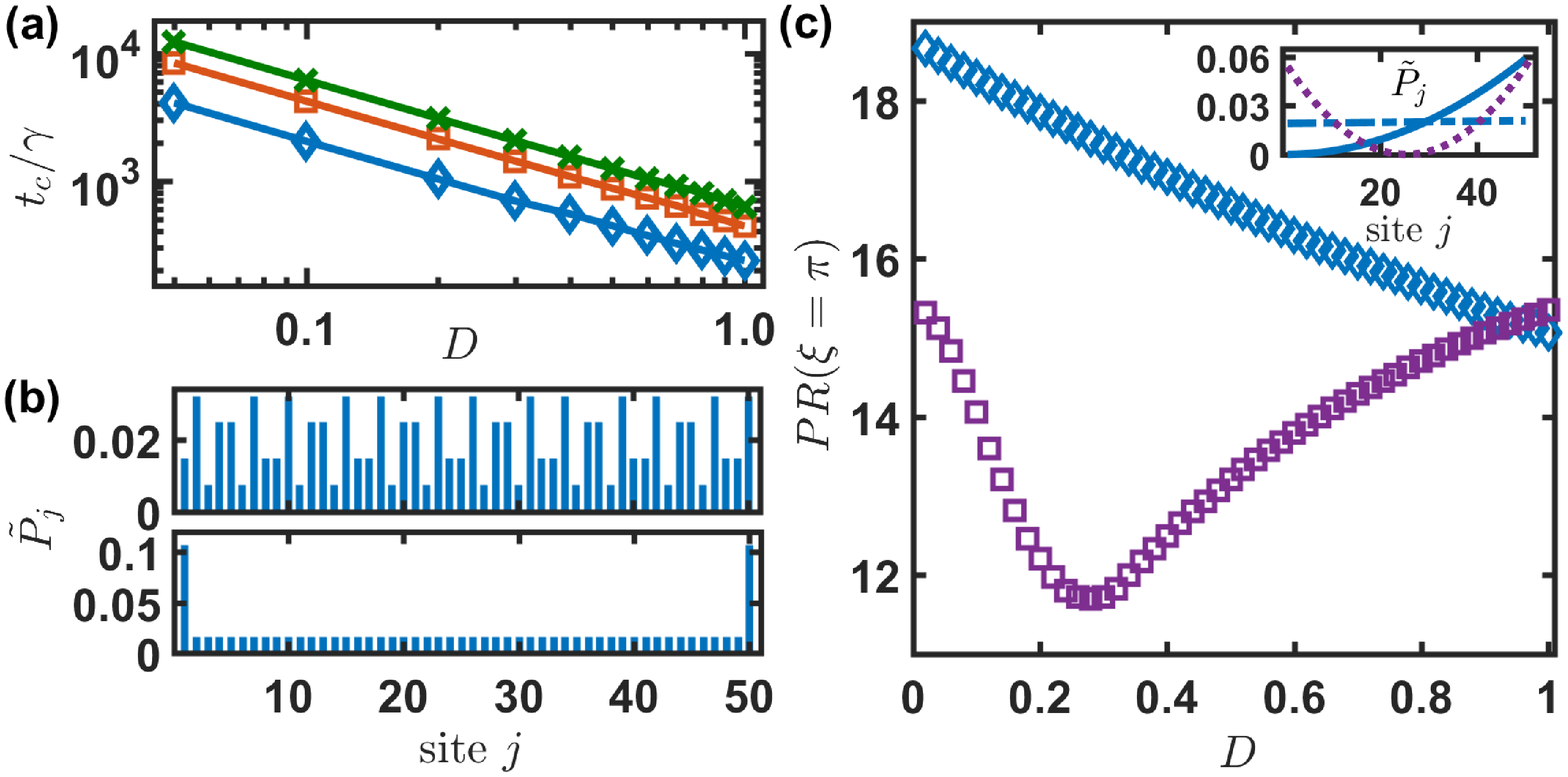}
\caption{Characteristics of critical points, CO, BEE, and CFD phases. (a) The time $t_c$ of atomic excitations in a log-log plot to go through the whole chain and reach the other end for various system sizes $N$ $=$ $50$ ($\diamond$), $100$ ($\square$), and $150$ ($\times$), at $\xi$ $=$ $0$. (b) A demonstration of states of CO ($D=1$, $\xi=\pi/4$) and BEE [($D=0$, $\xi=\pi/8$)] in the upper and lower panels respectively for $N=50$. (c) CFD phase shows two different $PR$ for $N$ $=$ $50$ ($\diamond$) and $51$ ($\square$), where state distributions $\tilde P_j$ in the inset present the cases of $D=1$ (solid) and $0.02$ (dash-dots) for $N$ $=$ $50$, and $D=0.02$ for $N=51$ (dots).}\label{fig2}
\end{figure}

We further locate two critical points at $D=0$, $\xi=0$ and $\pi$ respectively, and a phase of CFD which shows very different $PR$ for an even or odd number of atoms. The critically slow behavior can be identified by the time of atomic excitations to reach the other end, which has an algebraic dependence of directionality $D$ as shown in Fig. \ref{fig2}(a). This slow equilibrium of system dynamics also manifests in other extended phases. Criticality arises due to $(N-1)$ decoherence-free modes in the eigenvalues of $(-N/2,0,...,0)\gamma$, obtained from $M$, where zero decay modes are not allowed at other $\xi$'s. These decoherence-free modes mark a distinct region separating from other noncritical ones at $D$ $=$ $0$ with subradiant sectors of small but finite decay rates. 

We demonstrate two examples of CO and BEE phases in Fig. \ref{fig2}(b), where a period of eight sites and significant excitations at both edges emerge, respectively. CFD phase in Fig. \ref{fig2}(c) can be characterized by a dichotomy of steady states for even and odd number of atoms. For the first impression with a finite $D$, we expect of a flow of atomic excitations toward the preferred direction with its minimum in the opposite side. By contrast another configuration with its minimum moving toward the center of the chain as $D$ decreases shows up in odd number of atoms. It is the extra atom in an odd chain that distinguishes the two configurations near $D$ $=$ $0$ as shown in the inset of Fig. \ref{fig2}(c), which should satisfy the inversion symmetry as $D$ $\rightarrow$ $0$. Even(odd) number of atoms in this particular phase presents balanced(unbalanced) RDDI of alternating $\pm\gamma_{R/L}$, and this dichotomy can be further classified by two distinguishing $PR$'s. The more delocalized state for an even chain reaches its maximal $PR$ close to the critical point, whereas the $PR$ for an odd case never exceeds the value at $D$ $=$ $1$, and instead shows a minimum at a finite and universal value of $D$ $=$ $0.28$. The population accumulation toward both edges of the chain, representing a more localized phase, resembles an unpaired spin localization in a tight-binding Su-Schrieffer-Heeger model \cite{Su1979} or an edge state of bosons in a superlattice Bose-Hubbard model \cite{Grusdt2013}. We note that the cross of $PR$ near $D$ $=$ $1$ is due to finite size effect, which collapses in thermodynamic limit.

\begin{figure*}[hbtp]
\centering
  \begin{tabular}{c}
\includegraphics[width=16cm,height=7.5cm]{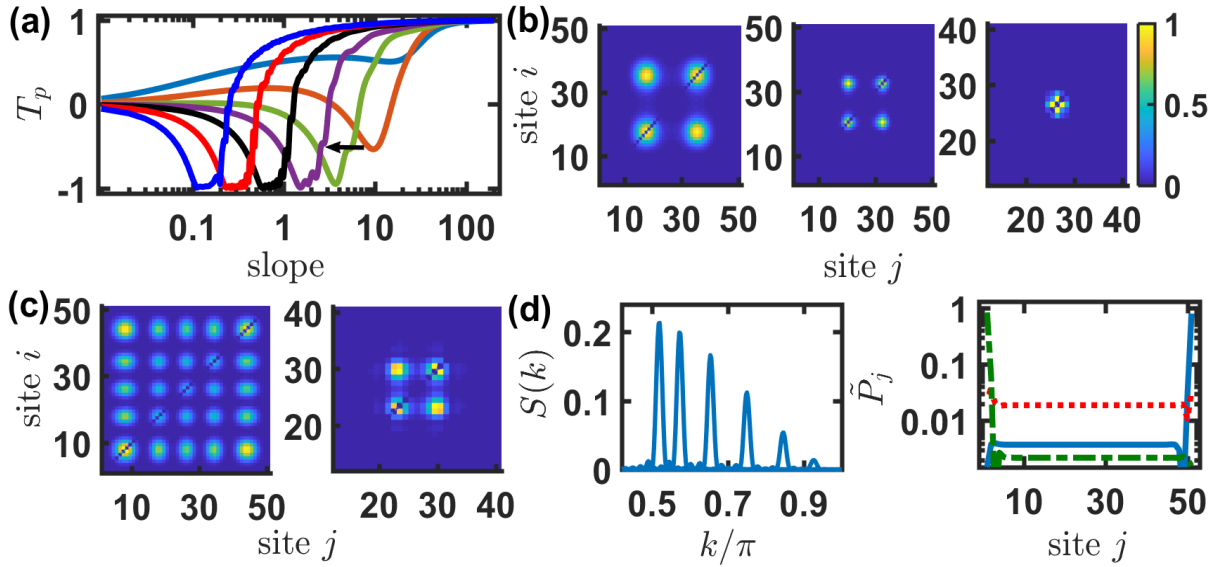}
\end{tabular}
\caption{Transport and localization of atomic excitations, emergence of CO, and the effect of $\theta_s$. (a) Interaction-driven transport of atomic excitations versus slopes(s) of linearly-increasing $\delta\mu$ $=$ $(s/N)(\mu-1)$. Arrows indicate the cases of $\xi$ $=$ $\pi/32$, $\pi/16$, $\pi/8$, $\pi/4$, $\pi/2$, $3\pi/4$, and $7\pi/8$, at $D$ $=$ $0$. (b) A localization of atomic excitations appears in atom-atom correlations $\tilde P_i\tilde P_j$ under a harmonic-potential-like $\delta_\mu$ $=$ $h[\mu-(N+1)/2]^2$ with $h$ $=$ $N^{-1}$ as $\xi$ increases from $\pi/4$ (left), $\pi/2$ (middle), to $3.9\pi/4$ (right), at $D$ $=$ $0$. (c) Edge accumulation is evident in $\tilde P_i\tilde P_j$ at $D$ $=$ $1$ with $\xi$ $=$ $3\pi/4$ in the left, and in the right broad CO emerge at $D$ $=$ $0.01$ and $\xi$ $=$ $\pi$, both under a smaller $h$ $=$ $0.01N^{-1}$ of harmonic-potential-like detunings with the same color bar in (b). (d) Finite structure factors $S(k)$ move from low to high $k$ from $\theta_s$ $=$ $7\pi/8$, $6\pi/8$, $...$, to $2\pi/8$, at $D$ $=$ $1$, $\xi$ $=$ $\pi/4$, and $N$ $=$ $101$. In the right panel, for $D$ $=$ $0.3$, $\xi$ $=$ $\pi/4$, the localized edge excitation moves from the right end at $\theta_s$ $=$ $7\pi/8$ (solid), to the left end at $\theta_s$ $=$ $\pi/8$ (dash-dots), comparing BEE phase at $\theta_s$ $=$ $\pi/2$ (dots). $N$ $=$ $51$ for (a-c) and right panel of (d).}\label{fig3}
\end{figure*}

{\it Transport of atomic excitations and emergence of crystalline order.--}Next we investigate the effect of spatially-dependent $\delta_\mu$ and $\theta_s$, which lead to redistribution of atomic excitations. We quantify the transport of atomic excitations by a difference of them between the left and right parts of the chain \cite{Jen2019_driven},  
\bea
T_p = \frac{\sum_{\mu=1}^{(N-1)/2}\tilde P_\mu - \sum_{\mu=(N+3)/2}^N \tilde P_\mu}{\sum_{\mu=1}^N \tilde P_\mu},
\eea
where we have excluded the central one for an odd $N$. Positive or negative $T_p$ represents that the left or right parts of the chain are more occupied, and right(left) linearly-increasing $\delta_\mu$ should favor positive(negative) $T_p$ since atoms are less excited under off-resonant driving fields, in a sense of noninteracting regime with negligible RDDI. By contrast in Fig. \ref{fig3}(a), for an onset of small slope of linearly-increasing detunings to the right, the interaction-driven transport changes a positive $T_p$ to negative one when $\xi$ $\gtrsim$ $\pi/8$, comparing $T_p$ $=$ $0$ for both BEE and sBHE phases at $D$ $=$ $0$ with a vanishing slope (symmetric distributions with inversion symmetry). This is more evident for moderate slopes, where negative $T_p$ shows up in different ranges, presenting the competition between RDDI strength and excitation detunings, that is, the case with a larger $\xi$ covers less ranges of slopes for negative $T_p$, indicating less effect of RDDI on transport properties. Eventually all $T_p$ turn to a positive side under large slopes as in noninteracting regimes. 

For transport of the excitations in BEE or sBHE phases, in Fig. \ref{fig3}(b) we show the atom-atom correlations which highlight the central localization of excitations as $\xi$ increases under a harmonic-potential-like detuning. This presents the repulsion of atomic excitations in BEE phase and the dominance of external potential over the hole excitations in sBHE phase. An example of edge accumulations induced from CO phase and finite CO emerged from CFD phase in Fig. \ref{fig3}(c) further shows the effect of $\delta_\mu$, which raises broadened CO. On the other hand, in Fig. \ref{fig3}(d) the effect of excitation angles $\theta_s$ manifests in moving the locations of finite structure factors by imprinting spatially-dependent phases on the chain, and in relocating the edge excitations to the left- or right-most of the chain as if a strong linear potential is induced for single-edge excitations. A joint manipulation of excitation detunings and angles thus enables a controllable transport of localized atomic excitations. 

\begin{figure}[t]
\centering
\includegraphics[width=8.5cm,height=4.5cm]{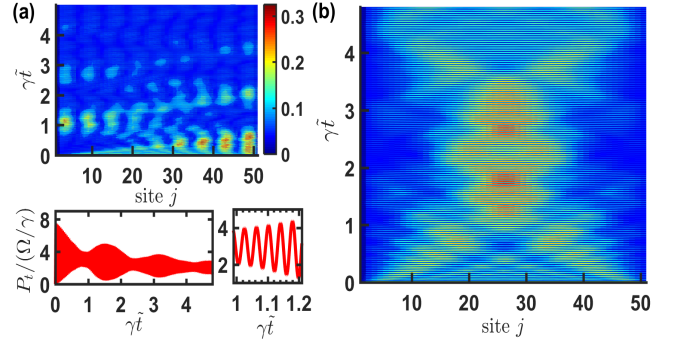}
\caption{Subharmonic oscillations and butterfly-like dynamics in the sBHE phase. (a) Time evolutions of $\tilde P_j$ and subharmonic oscillations of total populations $P_t$ $=$ $\sum_{\mu=1}^N\sigma_\mu^\dag\sigma_\mu$ (rescaled by $\Omega/\gamma$) in the upper and lower panels respectively, at $D$ $=$ $0.2$ and $\xi$ $=$ $0.8\pi$. (b) Butterfly-like time dynamics of $\tilde P_j$ at $D$ $=$ $0$ and $\xi$ $=$ $0.8\pi$ with the same color bar in (a). A rescaled time $\tilde t$ $=$ $1000 t$ represents the long time dynamics before reaching an equilibrium.}\label{fig4}
\end{figure}

{\it Persistent subharmonic oscillations.--}Here we present the long time dynamics of equilibration in steady states, specifically in the sBHE phase which shows strong extended features of hole excitations. There is no definite region for this collectively coupled dynamics, but can be approximately confined in mBHE and sBHE for $D$ $\lesssim$ $0.5$ and BEE for smaller $D$. 

In Fig. \ref{fig4}, we show two examples in non-cascaded ($D$ $\neq$ $0$) and reciprocal ($D$ $=$ $0$) coupling regimes. The non-cascaded time dynamics shows oscillating spread of populations from left to right initially since $\gamma_R$ $>$ $\gamma_L$, and then back and forth until equilibrium. We present an oscillation of total population $P_t$ with two time scales in the lower panel of Fig. \ref{fig4}(a), a signature of persistent subharmonic evolutions. This is further enhanced in the reciprocal coupling regime where much longer time is required to reach steady states. We attribute this phenomena as non-ergodic, associated to discrete time-crystalline order in subharmonic temporal responses \cite{Choi2017} and nonequilibrium many-body scars \cite{Turner2018} with a constrained local Hilbert space. The subradiant sectors of the eigen-spectrum under reciprocal couplings can provide some clues on this long time behaviors, where large energy splittings due to collective frequency shifts emerge toward the lowest subradiant decay rates as we go from BEE to sBHE phases along the line of $D$ $=$ $0$ \cite{SM}. This leads to highly dispersive couplings arising from strong RDDI and thus system dynamics that is far from equilibrium. The reoccurring patterns in Fig. \ref{fig4}(b) resemble a butterfly, where agglomeration of atomic excitations resides around the center of the chain before expanding to the edges, presenting another feature of non-ergodicity. As a final remark, we note of the scale of $P_t$ by $\Omega/\gamma$ instead of $(\Omega/\gamma)^2$, showing the excitation enhancement due to the collective couplings of RDDI, compared to the noninteracting case which should be order of $(\Omega/\gamma)^2$. 

In conclusion, the chiral-coupled atomic chain presents fruitful interaction-driven quantum phases of matter under driven-dissipative settings, with competing interactions between intrinsic 1D RDDI, directionality, and external excitation parameters. Under a weak excitation condition, the system shows critically slow behaviors, crystalline orders, and localized edge or extended hole excitations. The non-equilibrating states of extended hole excitations specifically manifest long time oscillations, in huge contrast to the states with CO and phases of BEE and wBHE close to the unidirectional coupling regime. Future directions can lead to unraveling clear mechanism for initiation of persistent subharmonic time evolutions, its relation to ergodicity of the chiral-coupled system, or many-body simulations of exotic or topological states in nonreciprocal quantum optical systems.

We acknowledge the support from the Ministry of Science and Technology (MOST), Taiwan, under the Grant No. MOST-106-2112-M-001-005-MY3 and thank Y.-C. Chen, G.-D. Lin, and M.-S. Chang for insightful discussions.

\clearpage
\begin{widetext}
\section{Supplemental Materials for Steady-state Phase Diagram of a Weakly Driven Chiral-coupled Atomic Chain}

\section{Hierarchy-coupled equations}
From a general model of driven-dissipative chiral-coupled atomic chain, 
\bea
\frac{d \rho}{dt}=-\frac{i}{\hbar}[H_S+H_L+H_R,\rho]+\mathcal{L}_L[ \rho]+\mathcal{L}_R[\rho],
\eea
we are able to derive the coupled equations with a hierarchy of multiple atomic correlations. We will see below that the hierarchy arises due to the resonant dipole-dipole interactions of $H_{L(R)}$ and $\mathcal{L}_{L(R)}[ \rho]$, where their explicit forms can be found in the main paper.

As a demonstration of hierarchy-coupled equations, we consider a uniform excitation field with Rabi frequency $\Omega$ at a right angle $\theta_s=\pi/2$. First, the time evolutions of coherence operators $\sigma_\mu$ reads \bea
\dot{\sigma}_\mu=\left(i\delta_\mu-\frac{\gamma_L+\gamma_R}{2}\right)\sigma_\mu+i\Omega(\sigma_\mu^{ee}-\sigma_\mu^{gg})-\gamma_L\sum_{\nu>\mu}e^{ik_s|r_{\mu,\nu}|}(\sigma_\mu^{gg}-\sigma_\mu^{ee})\sigma_\nu-\gamma_R\sum_{\nu<\mu}e^{ik_s|r_{\mu,\nu}|}(\sigma_\mu^{gg}-\sigma_\mu^{ee})\sigma_\nu,\label{dipole}
\eea
where $\gamma_{L(R)}$ respectively quantifies the couplings between the atom to the rest of left(right) of the chain. Next, the evolution of excitation population reads
\bea
\dot{\sigma}_\mu^{ee}=i\Omega\sigma_\mu-i\Omega\sigma_\mu^\dag-\gamma_L\sum_{\nu>\mu}\left(e^{ik|r_{\mu,\nu}|}\sigma_\mu^\dag\sigma_\nu+H.c.\right)-\gamma_R\sum_{\nu<\mu}\left(e^{ik|r_{\mu,\nu}|}\sigma_\mu^\dag\sigma_\nu+H.c.\right),
\eea
where $H.c$ is Hermitian conjugate. We can see that the above two single atomic operators couple with each other via two-body correlations of $\sigma^{gg}_\mu\sigma_\nu$, $\sigma^{ee}_\mu\sigma_\nu$, and $\sigma^\dag_\mu\sigma_\nu$. This indicates the hierarchy relations between $n$th and $(n+1)$th moments of operators up to $N$th ones. 

To have another taste of hierarchy-coupled equation of two-body correlations, we show the time evolutions of $\sigma_\mu^\dag\sigma_\nu$ as an example, which reads 
\bea
\frac{d(\sigma_\mu^\dag\sigma_\nu)}{dt}=&&-(\gamma_R+\gamma_L)\sigma_\mu^\dag\sigma_\nu-i\Omega[(\sigma_\mu^{ee}-\sigma_\mu^{gg})\sigma_\nu-\sigma_\mu^\dag(\sigma_\nu^{ee}-\sigma_\nu^{gg})]\nonumber\\
&&-\gamma_L\left[\sum_{\alpha>\nu}e^{ik|r_{\alpha,\nu}|}\sigma_\mu^\dag(\sigma_\nu^{gg}-\sigma_\nu^{ee})\sigma_\alpha+\sum_{\alpha>\mu}e^{-ik|r_{\alpha,\mu}|}(\sigma_\mu^{gg}-\sigma_\mu^{ee})\sigma_\alpha^\dag\sigma_\nu\right]\nonumber\\
&&-\gamma_R\left[\sum_{\alpha<\nu}e^{ik|r_{\alpha,\nu}|}\sigma_\mu^\dag(\sigma_\nu^{gg}-\sigma_\nu^{ee})\sigma_\alpha+\sum_{\alpha<\mu}e^{-ik|r_{\alpha,\mu}|}(\sigma_\mu^{gg}-\sigma_\mu^{ee})\sigma_\alpha^\dag\sigma_\nu\right],
\eea
where three-body operators emerge. In treating the effect of quantum fluctuations in dissipative quantum optical systems \cite{Carmichael2008}, we can usually truncate the hierarchy up to two-body noise correlations due to the stochastic nature of quantum noises. For a general driven-dissipative system we consider here, many-body states can be all explored (a total of $2^N$ configurations for $N$ two-level quantum registers), and this is exactly the merit of quantum resource for quantum information processing and quantum computation. However, this also prevents appropriate numerical simulations of general quantum dynamics by classical computers, except for systems with smaller higher order moments or with a technique of matrix product states which are efficient in finding the many-body ground states.

Next we simplify the hierarchy-coupled equations by taking a low saturation regime, where $\sigma_\mu^{gg}\approx 1\gg\sigma_\mu^{ee}$. Under this condition, Eq. (\ref{dipole}) reduces to
\bea
\dot{\sigma}_\mu=\left(i\delta_\mu-\frac{\gamma_L+\gamma_R}{2}\right)\sigma_\mu-i\Omega-\gamma_L\sum_{\nu>\mu}e^{ik_s|r_{\mu,\nu}|}\sigma_\nu-\gamma_R\sum_{\nu<\mu}e^{ik_s|r_{\mu,\nu}|}\sigma_\nu,\label{dipole2}
\eea
and the evolutions of all other higher order moments can be decomposed in terms of the above one. For example, $\langle\sigma_\mu^\dag\sigma_\nu\rangle=\langle\sigma_\mu^\dag\rangle\langle\sigma_\nu\rangle$, where $\langle\cdot\rangle$ denotes the expectation values, such that $d\langle\sigma_\mu^\dag\sigma_\nu\rangle/dt=\langle\dot\sigma_\mu^\dag\rangle\langle\sigma_\nu\rangle+\langle\sigma_\mu^\dag\rangle\langle\dot\sigma_\nu\rangle$. In the main paper, we solve the steady-state solutions of Eq. (\ref{dipole2}), which manifest many interesting quantum phases in the steady state.

\section{Finite structure factors in thermodynamic limit}
To identify crystalline orders (CO), we use structure factors defined as
\bea
S(k)=\sum_{j=1}^N \sum_{m=1}^Ne^{ik(j-m)}\tilde P_j\tilde P_m,
\eea
to characterize the phase with CO. $\tilde P_j$ denotes the normalized steady-state population,r
\bea
\tilde P_j= \frac{\langle\sigma_{j}^{(s)\dag}\sigma_{j}^{(s)}\rangle}{\sum_{j=1}^N \langle\sigma_j^{(s)\dag}\sigma_j^{(s)}\rangle}.
\eea

The phase boundaries of CO-BEE (bi-edge excitations) and CO-BHE (bi-hole excitations) in the main paper are determined respectively by abrupt occurrences of edge or hole excitations at both ends of the chain. This coincides with clear differences of participation ratios between extended and localized states, and with a distinction by a finite CO. The ETD (extended distributions)-CO phase boundary, however, is not obvious since ETD and CO phases are both extended. Close to this phase boundary, the structure factors are decreasing as $N$ increases. Therefore, we fit the maximum (Max.) of $S(k\neq 0)$ in terms of $N$ by $(aN^{-1}+b)$ with fitting parameters $a$ and $b$, where a finite $b>0$ represents a finite $S(k)$ in thermodynamic limit of $N\rightarrow\infty$, and we characterize it as CO phase. 

We delineate the ETD-CO phase boundary by fitting four pints of $N=100-400$ in the phase diagram. As an example, we show the case at $D=0.05$ and extract the Max. of structure factors at $k\neq 0$,
\begin{center}
    \begin{tabular}{ | l | l | l |}
    \hline
    N & Max. of $S(k)$ at $\xi=0.001$ & Max. of $S(k)$ at $\xi=0.002$\\ \hline
    100 & 4.4{\rm E}-4  & 3.58{\rm E}-4 \\ \hline
    200 & 9{\rm E}-5    & 6.26{\rm E}-4 \\ \hline
    300 & 1.09{\rm E}-4 & 3.74{\rm E}-4 \\\hline
		400 & 1.56{\rm E}-4 & 3.42{\rm E}-4 \\
    \hline
    \end{tabular}
\end{center}
where fitting parameters of $(a,b)$=(0.043,-3{\rm E}-5) and (-0.0013,4{\rm E}-4) respectively, making $\xi=0.002$ a point on the phase boundary separating ETD and CO phases. Since this boundary is very close to the line of $\xi=0$, we specify the boundary with a precision up to $\xi=\pm 0.001$. From the above table, we note of the oscillating $S(k)$ as $N$ increases, and thus more points for fitting makes no significant changes in this estimate of the phase boundary. 

\section{Phase boundaries from BEE to BHE and between sBHE, mBHE, and wBHE}
\subsection{Phase boundaries from BEE to BHE}

Here we show the analytical results for the phase boundary at $\xi=\pi/3$ from BEE to BHE at $D=0$, and the numerically obtained boundaries between sBHE, mBHE, and wBHE. For $\delta_\mu=0$ and $\theta_s=\pi/2$ considered in the phase diagram in the main paper, we have the coupling matrix at $D=0$,
\bea
M=
\begin{bmatrix}
-\frac{1}{2} & -\frac{1}{2}e^{i\xi} & -\frac{1}{2}e^{i2\xi} & ... & -\frac{1}{2}e^{i(N-1)\xi} \\
-\frac{1}{2}e^{i\xi} & -\frac{1}{2} & -\frac{1}{2}e^{i\xi} & ... & -\frac{1}{2}e^{i(N-2)\xi} \\
-\frac{1}{2}e^{i2\xi} & -\frac{1}{2}e^{i\xi} & -\frac{1}{2} & ... & \vdots\\
\vdots & \vdots & \hdots & \ddots & \vdots \\
-\frac{1}{2}e^{i(N-1)\xi} & -\frac{1}{2}e^{i(N-2)\xi} & ... & ... & -\frac{1}{2}
\end{bmatrix},
\eea
which is symmetric, and so is its inverse $M^{-1}$. At the BEE-BHE phase boundary, $\tilde P_{1(N)}$ should be equal to $N^{-1}$ to distinguish from edge ($\tilde P_{1(N)}>N^{-1}$) and hole ($\tilde P_{1(N)}<N^{-1}$) excitations. 

We can obtain $\tilde P_j$ from $M^{-1}$. For $N=3$, and take $\tilde P_1$ ($\tilde P_N=\tilde P_1$) as an example,
\bea
\tilde P_1= \frac{A}{2A+A^2},~ A\equiv |1-e^{i\xi}|^2,
\eea
which gives $\xi=\pi/3$ for $A=1$ when $\tilde P_1=1/3$. Another solution of $\xi\approx 0$ from $A=0$ is ignored due to the divergence of $\tilde P_1$. This relates to critical regimes, which we will discuss later. For a general $N$, we obtain the solution of $\xi$ from 
\bea
\tilde P_1= \frac{A}{2A+(N-2)A^2}=\frac{1}{N},
\eea
which gives again $\xi=\pi/3$ for $A=1$. When $\xi\approx 0$, we obtain $\tilde P_{1(N)}=1/2$, which indicates that the atomic excitations are populated equally to the edges of the chain. When $\xi=\pi$, we have $\tilde P_{1(N)}=1/(4N)$ for $N\gg 1$, a signature of hole or null excitations at the edges.  

\subsection{Phase boundaries between sBHE, mBHE, and wBHE}

In the BHE phase, we note of the suppression of the participation ratios (PR) as $D$ increases, which indicates that BHE becomes less delocalized. To quantify various regions of PR, we again fit the maximum of PR in terms of $N$ by $\beta N^{\alpha}$ with fitting parameters $\beta$ and $\alpha$, where $\alpha$ denotes the degree of scalings for strong (s), moderate (m), and weak (w) regimes with $\alpha>0.5$ (below $D\approx 0.37$), $0.1<\alpha<0.5$, and $\alpha<0.1$ (above $D\approx 0.67$), respectively. This algebraic dependence of $N$ is suggested at $D=0$, where PR$\propto N^{-1}$.

Below we show the PR and their fitting parameter of $\alpha$ around sBHE, mBHE, and wBHE regimes by three data points, 
\begin{center}
    \begin{tabular}{ | l | l | l | l | l |}
    \hline
    N & PR at $D=0.37$ & PR at $D=0.38$ & PR at $D=0.67$ & PR at $D=0.68$ \\ \hline
    25  & 13.61 & 13.15 & 3.47 & 3.32 \\ \hline
    50  & 20.89 & 19.87 & 3.82 & 3.61 \\ \hline
    100 & 27.65 & 25.92 & 4.012& 3.77 \\ \hline \hline
		$\alpha$ & 0.51 & 0.49& 0.105& 0.092 \\ \hline
    \end{tabular}
\end{center}

\section{Subradiant sectors of the eigen-spectrum}

\begin{figure}[b]
\centering
\includegraphics[width=16cm,height=8cm]{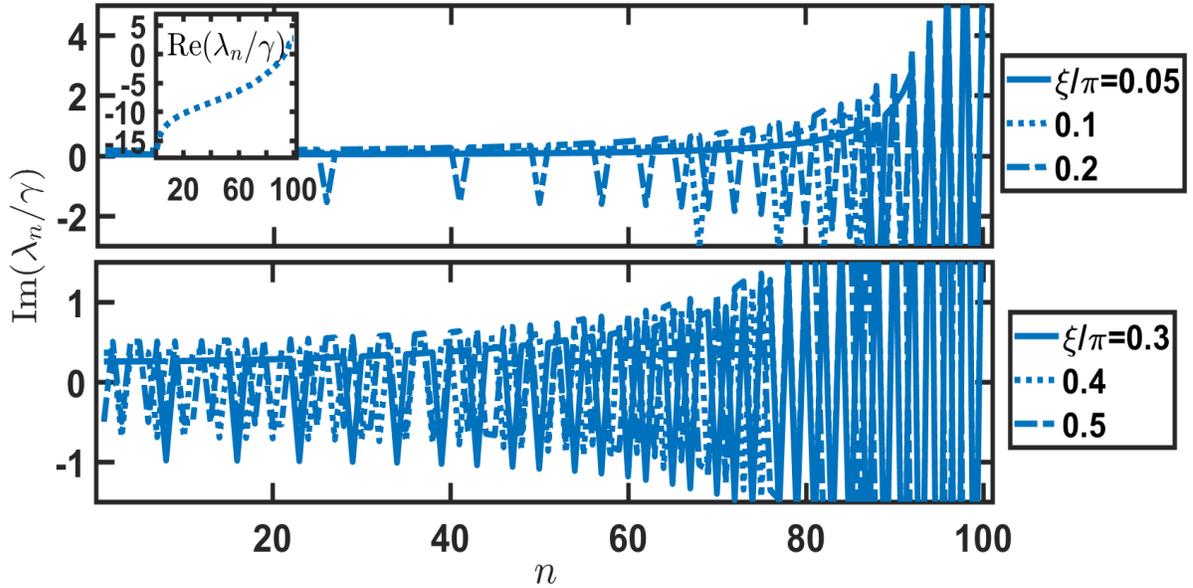}
\caption{Frequency shifts of the eigen-spectrum for $N=100$. Here we sort the eigenvalues Im($\lambda_n$) according to ascendant decay rates of Re($\lambda_n$). In the upper panel, the inset shows the case of $\xi/\pi=0.1$ in logarithmic scales. We specifically focus on the subradiant sectors of $n\lesssim 80$, which is way below Re($-\lambda_n$) $=$ $\gamma$.}\label{supfig1}
\end{figure}

Here we take $N=100$ as an example, and show below the energy splittings of the eigen-spectrum from BEE to sBHE in the phase diagram at $D=0$. The eigen-spectrum is directly obtained from the coupling matrix $M$ defined in the main paper. In Fig. \ref{supfig1}, we sort the frequency shifts of $\lambda_n$ according to the sorting from low to high decay rates as $n$ increases. The trend of the logarithmic scales of eigen decay rates is similar for all cases of $\xi$, except some larger superradiant or lower subradiant eigenvalues appear for different parameters of $\xi$. We scan $\xi$ from BEE ($\xi<\pi/3$) to sBHE ($\xi>\pi/3$), and find the emergence of energy splittings appearing toward the lowest subradiant sectors. Since the long time behaviors of subharmonic oscillations observed in the main paper are more significant in the phase of sBHE, we attribute the non-ergodic oscillations to the highly dispersive subradiant sectors of the spectrum.

This can be further clarified in Fig. \ref{supfig2}, where we plot the frequency shifts in an ascendant direction. Again as we go from BEE to sBHE phases, the eigen frequency shifts close to Im($\lambda_n$) $=$ $0$ start to split, and an energy gap-like jump appears from red-detuned to blue-detuned shifts. This opening of energy gap corresponds to the subradiant sectors which allow relatively lower decay rates with finite energy shifts. In the perspective of steady states under driven-dissipative settings, these constrained subradiant states are responsible for the persistent subharmonic oscillations or butterfly-like time dynamics we present in the main paper.

\begin{figure}[t]
\centering
\includegraphics[width=16cm,height=8cm]{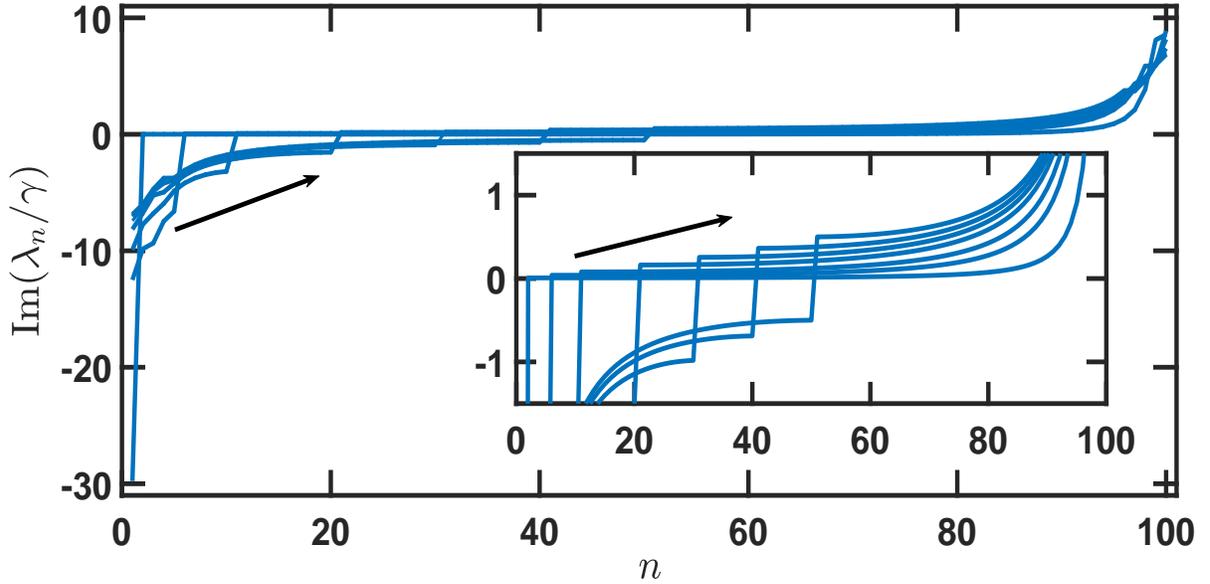}
\caption{Frequency shifts of the eigen-spectrum for $N=100$ from low to high values as $n$ increases. The inset zooms in the values near Im($\lambda_n$) $=$ $0$, and the arrows marks the direction of increasing $\xi/\pi=0.01$, $0.05$, $0.1$-$0.5$.}\label{supfig2}
\end{figure}
\end{widetext}
\end{document}